# Letter to the Editor Concerning "Simultaneous, Single-Particle Measurements of Size and Loading Give Insights into the Structure of Drug-Delivery Nanoparticles"

Andrew C. Madison,‡ Adam L. Pintar,‡ Craig R. Copeland, Natalia Farkas, and Samuel M. Stavis*

The vexing error of excess variance in measurements of nanoparticle size distributions degrades accuracy in applications ranging from quality control of nanoparticle products to hazard assessment of nanoplastic byproducts. The particular importance of lipid nanoparticles for vaccine and medicine delivery motivates this Letter to the Editor, which concerns a publication[1] in *ACS Nano* and also presents original research. In ref 1, the benchmark measurements of a nanoparticle standard manifest large errors of the size distribution that contradict the claim of validation. In subsequent applications of the method to measure lipid nanoparticles, potential errors could bias the correlation of fluorescence intensity as an optical proxy for molecular loading and give misleading insights from power-law models of intensity–size trends. Looking forward, measurement error models may address this issue.

A brief review provides helpful context. Heterogeneous sizes and properties of colloidal nanoparticles are challenging to measure. Measurements of nanoparticle ensembles often cannot resolve size and property distributions, providing only estimates of means—ordinarily arithmetic means but occasionally harmonic means—and variances. For example, photon correlation spectroscopy or dynamic light scattering,[2] which is in common use and abuse for measuring these values,[3] as well as fluorescence correlation spectroscopy,[4] which contends with focal volume artifacts,[5] rearrange the Stokes–Einstein–Sutherland[6,7] equation to infer hydrodynamic diameter from ensemble diffusivity. Ensemble measurements can output some information about the molecular loading of lipid nanoparticles if size distributions are available as additional inputs,[8] but measurements of single nanoparticles can be more informative and less dependent on model assumptions. Nanoparticle tracking analysis,[9] which is the basis of the method in Ref 1, infers the hydrodynamic diameters of single nanoparticles from optical microscopy of diffusive trajectories, and also yields estimates of optical properties such as fluorescence intensity for correlation. However, short trajectories through the focal volume of an optical microscope limit sizing resolution, motivating Bayesian statistical analyses to improve accuracy[10] and fluidic devices to confine nanoparticles, prolong tracking, and improve sizing resolution.[11]

**Measurement function.** A consequence of fluidic confinement is a reduction of nanoparticle diffusivity through hydrodynamic interactions, requiring a correction for this systematic effect. In the simplest case of a confining slit, the resulting measurement function is

$$d_\mathrm{h} = \frac{k_\mathrm{B}}{3\pi} \cdot \frac{T}{\eta} \cdot \frac{C}{D_{||}} \qquad (1)$$

where $d_\mathrm{h}$ is the hydrodynamic diameter of a single nanoparticle, $k_\mathrm{B}$ is the Boltzmann constant, $T$ is the absolute temperature of the dispersion fluid, $\eta$ is the kinematic viscosity of the dispersion fluid, $C$ is a hydrodynamic correction, and $D_{||}$ is the diffusivity parallel to the slit surfaces (Table S1). In ref 1

$$C \approx \left(1 - 1.004\left(\frac{d_\mathrm{h}}{d_\mathrm{s}}\right) + 0.418\left(\frac{d_\mathrm{h}}{d_\mathrm{s}}\right)^3 + 0.210\left(\frac{d_\mathrm{h}}{d_\mathrm{s}}\right)^4 - 0.169\left(\frac{d_\mathrm{h}}{d_\mathrm{s}}\right)^5\right) \qquad (2)$$

where $d_\mathrm{s}$ is the slit depth.[12,13]

**Input quantities.** The measurement function in eq 1 has multiple input quantities that require consideration and that can introduce systematic effects, whereas ref 1 reports only nominal values of $d_\mathrm{s}$ and apparent values of $D_{||}$. Regarding $T/\eta$, quantitation of fluid temperature and viscosity, and validation of temperature stability and fluid quiescence, are all necessary but absent from ref 1. Regarding $C/D_{||}$, the hydrodynamic correction[12,13] applies only at the midplane of a slit and deviates significantly from the results of careful measurements of microparticle diffusion near the surface of a slit,[14] and a mean correction neglects the varying diameters of single nanoparticles. Further, the analysis in ref 1 extracts $D_{||}$ from a model of diffusion in two dimensions in a confining circle,[15] whereas $C$ corrects the parallel component of diffusion in three dimensions in a confining slit. Neither aspect matches the experimental scheme of diffusion in a confining cylinder, and a representative model fit systematically underestimates the experimental data at short lag times (Figure 1D of ref 1). As well, optical microscopy throughout a wide[16] and deep[17] focal volume requires calibration of scale factor and aberration artifacts such as distortion and defocus to correct apparent trajectories that are the source data for $D_{||}$. Finally, a formal measurement requires accurate estimates of not only input quantities but also input uncertainties, and their propagation through the measurement function to an output quantity and output uncertainty,[18] which is incomplete in ref 1.

**Table 1. Data from Table 2 of Reference 1 and Figure 1F of Reference 1.**

| property | source | quantity (nm) | | | | |
|---|---|---|---|---|---|---|
| **Microwell depth** | **Table 2** | **200** | **350** | **500** | **800** | **1200** |
| **Particle size from mean diffusivity** | **Table 2** | **62 ± 2** | **45 ± 2** | **47 ± 1** | **48 ± 2** | **43 ± 1** |
| Harmonic mean diameter | Figure 1F | 61.3 ± 2.5 | 46.8 ± 1.0 | 48.6 ± 0.7 | 48.2 ± 0.8 | 45.3 ± 1.2 |
| Arithmetic mean diameter | Figure 1F | 73.1 ± 3.2 | 52.9 ± 1.0 | 52.8 ± 0.8 | 52.2 ± 0.8 | 50.8 ± 1.3 |
| Standard deviation of diameter | Figure 1F | 33.9 ± 3.6 | 18.6 ± 0.7 | 15.9 ± 0.6 | 14.7 ± 0.6 | 18.3 ± 1.0 |

Bold text indicates data reported in Table 2 of Reference 1. Black text indicates new analysis of Figure 1F of Reference 1. Uncertainties therein are 68 % coverage intervals of digitizing errors and sampling variability (Supporting Information, Notes S1-S3 of this Letter).

**Nanoparticle standard.** Even after a complete evaluation of uncertainty, independent validation is necessary. The effort to validate in ref 1 involves tests of a nanoparticle standard, consisting of polystyrene spheres loaded with hydrophobic fluorophores, in cylindrical microwells of five depths (Table 1 of this Letter). The choice of the standard nominally matches the sizes of lipid nanoparticles in ref 1. Validation would require agreement of the test results with at least two reference values of the diameter distribution—a mean of 47.8 ± 0.3 nm and a root variance or standard deviation of 6.3 ± 0.2 nm.[19] These uncertainties are 68 % coverage intervals of sampling variability only, neglecting potential uncertainty from other sources. Moreover, the experimental method of ref 1 yields a hydrodynamic size, diverging from the reference method of transmission electron microscopy,[19, 20] which yields a steric size. Setting aside this divergence, this key comparison suffices to critically evaluate the foundational claim of ref 1—validation of the measurement method by agreement with the expected size within ± 1 nm.

**Fitting analysis.** Reference 1 involves sizing of single nanoparticles but also aspects of ensemble analysis. Reference 1 claims that the histograms in Figure 1F show the collapse of different diffusivity distributions onto the same size distribution after hydrodynamic correction, excluding data from the shallowest device. Yet, the analysis in ref 1 calculates the mean diffusivity for each microwell depth and fits the hydrodynamic correction to all of the data, including data from the shallowest device, reporting a peak diameter of 49 ± 6 nm. This value of fit uncertainty is distinct from the reference value of standard deviation—the two values are only coincidentally equal, as the following subsection on distributional errors shows. The fitting analysis in ref 1, or alternatively combining the diameter histograms after hydrodynamic correction, could cause random effects to cancel, as would occur for replicate measurements. However, several of the systematic effects depend on microwell depth, and an assumption that they should cancel would be questionable.

**Independent analysis.** This issue motivates a new analysis for each microwell depth independently (Table 1 of this Letter), which assumes nothing of the correctness of fitting or combining all of the data, and yields two root-mean-square errors as conventional metrics of accuracy. This analysis involves data from all five microwell depths, matching the effort to validate, as well as one of the three applications of the method to lipid nanoparticles in ref 1. The analysis begins with calculation of the harmonic mean, arithmetic mean, and standard deviation of each diameter histogram in Figure 1F of ref 1 (Supporting Information, Tables S2-S7 and Notes S1-S3 of this Letter).[21, 22] The nanoparticle sizes from mean diffusivity in Table 2 of ref 1 and the harmonic mean diameters agree within uncertainty, which is consistent with the inverse proportionality of arithmetic mean diffusivity and harmonic mean diameter in the Stokes–Einstein–Sutherland equation. A similar effect results from ensemble averaging in dynamic light scattering,[3] building confidence in the new analysis. However, the analysis in ref 1 overlooks this relation, resulting in a critical inconsistency—the key comparison to the reference values requires the arithmetic means, rather than the harmonic means, as well as the standard deviations of the diameter distributions.

**Distributional errors.** All of the arithmetic mean diameters exceed the reference value of 47.8 ± 0.3 nm (Figure 1a of this Letter), yielding a root-mean-square error of 12.0 nm or 25.1%. This error is an order of magnitude greater than the corresponding claim of ref 1, confirming the presence of systematic effects. The standard deviations of diameter are even more problematic. Reference 1 acknowledges excess variance due to short trajectories from fluorescence photobleaching but avoids quantifying the distribution widths. All of the standard deviations of diameter exceed the reference value of 6.3 ± 0.2 nm (Figure 1a of this Letter), yielding a root-mean square error of 15.6 nm or 247%. This large error corresponds to diameter histograms that are several times wider than the reference diameter distribution, showing low resolution for sizing single nanoparticles. On this basis, the claim of validation of the measurement method by agreement with the expected size within ± 1 nm lacks substantiation.

**Experimental disconnect.** In the subsequent applications of the method in ref 1, a different objective lens and imaging conditions prolong the tracking analysis from 600 to 2000 images. Although longer trajectories could reduce the error of the standard deviation[11] by up to a factor of $\sqrt{2000/600} = 1.83 \approx 2$, different systematic effects could also be present due to scale factor and aberration artifacts intrinsic to objective lenses.[16, 17] As well, ref 1 notes potential errors due to a hard sphere analysis of soft lipid nanoparticles with complex diffusive behavior. This mismatch of experimental conditions and potential errors disconnects the efforts to validate and apply the method.

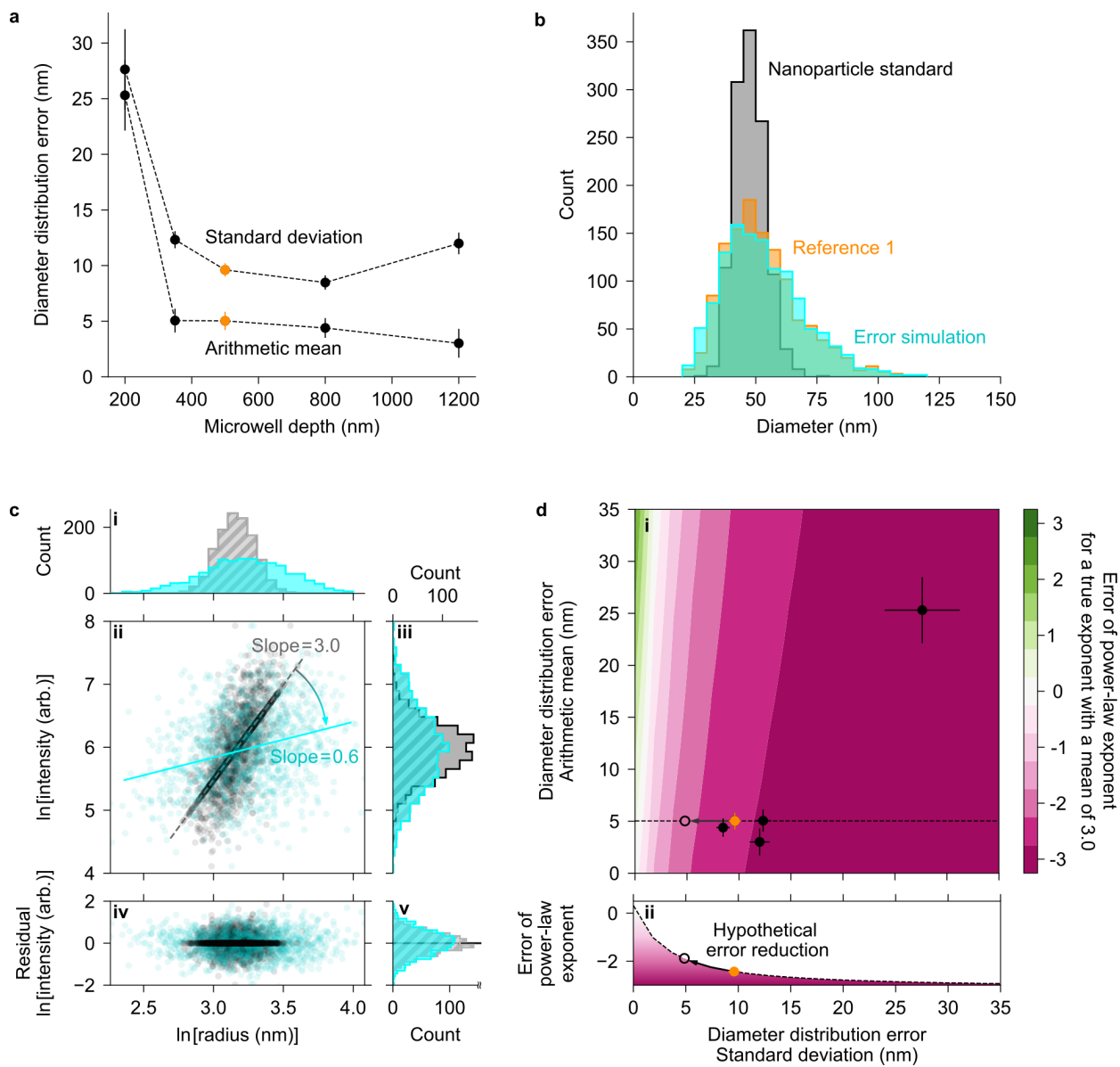

Figure 1. (a) Scatter plot showing diameter distribution errors from the benchmark measurements of a nanoparticle standard in Reference 1. Vertical bars are 68 % coverage intervals of digitizing errors and sampling variability (Supporting Information of this Letter). (b) Histograms showing representative diameter distributions of (gray) the nanoparticle standard, (orange) Reference 1, for the corresponding orange data marker in (a), and (cyan) an error simulation. (c) Marginal histograms and scatter plots showing a representative negative bias that is evident as a slope suppression in a linear model of ln(intensity) *versus* ln(radius). (c-i, iii) (Solid gray with black outline) Marginal histograms showing reference data without measurement errors. (Hatch gray with gray outline) Marginal histograms showing data with ln(intensity) errors only. (Solid cyan) Marginal histograms showing data with both ln(intensity) and ln(radius) errors. (c-ii) Scatter plots showing ln(intensity) *versus* ln(radius) data (black data markers) without measurement errors, (gray data markers) with ln(intensity) errors, and (cyan data markers) with both ln(intensity) and ln(radius) errors. (Dash gray) Line showing least-squares fit to gray data markers. (Solid cyan) Line showing least-squares fit to cyan data markers. (c-i, v) Scatter plot and marginal histograms showing residual values of ln(intensity) from the fits in (c-ii), with approximately normal distributions around means of zero. (d-i, ii) Contour plot and (dash section) line plot showing errors of the power-law exponent for a true exponent with a mean of 3.0 in a factorial analysis as a function of diameter distribution errors. The five data markers in (d-i) correspond to the five microwell depths in (a). Uncertainty of the error of the power-law exponent in (d-ii) is smaller than the orange data marker. The black arrow and open circle correspond to a hypothetical halving of a representative error of the standard deviation between the efforts to validate and apply the method.

**Potential bias.** This disconnect results in uncertain sizing errors in applications of the method. However, even allowing for a hypothetical halving of the error of the standard deviation, a potential bias[23] raises a question about the reliability of the results. A major goal of ref 1 is to use power-law models of the dependence of fluorescence intensity on hydrodynamic size to study the loading of fluorophores into lipid nanoparticles. In ordinary least-squares fits of such data, unrecognized errors of the independent variable of size can bias estimated parameters of models of the dependent variable of intensity.[23] The potential bias depends on the widths of both the true and measurement error distributions of the independent variable[24] and becomes critical in comparisons of expected and observed power-law exponents, which guide interpretation of the results in ref 1. Reference 1 omits such analysis for the standard, which presents another opportunity to test expectations against observations, given sufficient sizing resolution.[25, 26]

**Size simulations.** Monte-Carlo simulations elucidate the potential bias for the reference diameter and sizing error distributions in ref 1 (Supporting Information, Notes S4 of this Letter), whereas both distributions are uncertain for the lipid nanoparticles. Without access to a reference diameter histogram for the nanoparticle standard, the reference mean and standard deviation parameterize a lognormal distribution of diameter, $d$. This common model allows characteristic asymmetry of nanoparticle size distributions.[27, 28] A symmetric normal distribution yields similar results (Supporting Information of this Letter). Skew-normal distributions of diameter errors approximate unknown combinations of random and systematic effects in ref 1, with model parameters resulting from deconvolution of the experimental histograms in Figure 1 of ref 1 with the reference diameter distribution. This new analysis yields synthetic histograms that closely match the experimental histograms in Figure 1F of ref 1 (Figure 1b, Supporting Information, Figure S1 of this Letter). Per the choice of the standard, these histograms at least resemble if not match the diameter histograms of lipid nanoparticles in Figures 2, 3, and 6 of ref 1.

**Intensity simulations.** A power law then constrains fluorescence intensity, $I$, by the model, $I = \gamma r^\beta$, with a coefficient, $\gamma$, an exponent, $\beta$, and changing the independent variable from diameter $d$ to radius $r$, as in ref 1. A transformation of this model by the natural logarithm, $\ln(I) = \beta \ln(r) + \ln(\gamma)$, serves three purposes. First, this transformation facilitates modeling of $\ln(\text{intensity})$ errors by a normal distribution with a mean of zero and a range of standard deviations that approximate the vertical scatter of $\ln(\text{intensity})$ values at the central radii in Figures 2, 3, and 6 of ref 1, capturing both measurement variability and sample heterogeneity (Figure 1c, Supporting Information, Figure S2, and Table S8 of this Letter). Second, this transformation enables ordinary least-squares fits of linear models to the $\ln(\text{intensity})$ *versus* $\ln(\text{radius})$ data, with the slope $\beta$ equaling the power-law exponent, as in Figure 6 of ref 1. Third, this transformation facilitates a comparison to the canonical example of slope suppression that motivates a measurement error model.[23]

**Slope suppression.** A factorial study shows that the predominant effect is a suppression of the slope by the error of the standard deviation of the diameter distribution (Figure 1c, Supporting Information, Figures S2 and S3 of this Letter). This study considers true slopes ranging from 1.5 to 3.5, with a value of 3.0 matching the upper limit expected in ref 1. This negative bias has an intuitive explanation—excess variance increases the run of $\ln(\text{radius})$ for a constant rise of $\ln(\text{intensity})$, reducing the slope of the trend. This slope, which is equivalent to the exponent of a power-law model, nominally gives physical or chemical insight, such as by the proportionality of fluorophore count to loading processes that depend on area or volume. Accordingly, this bias of the power-law exponent can give misleading insights, potentially confounding intensity–size correlation in ref 1. For the range of distributional errors in the effort to validate in ref 1, errors of the arithmetic mean that are much larger than errors of the standard deviation can cancel this bias (Figure 1d of this Letter). However, the distributional errors in the effort to validate have the opposite relation, indicating that the results lie outside of this cancellation band, even allowing for a hypothetical halving of the error of the standard deviation between efforts to validate and apply the method (Figure 1d of this Letter). Moreover, even if a cancellation of errors were to occur in subsequent applications of the method, this would constitute an unreliable basis for an accurate measurement, substantiating the question of reliability.

**Conclusion.** The scope of ref 1 is ambitious, the experiments are challenging, and the effort of the authors is commendable. However, a distributional validation is necessary to fairly compare and reliably apply methods to measure single nanoparticles. Agreement of both the mean and the standard deviation of a nanoparticle size distribution near the scale of 1 nm requires comprehensive calibration and reveals the potential for unexpectedly high power-law exponents modeling intensity–size trends, even for the common standards benchmarked in ref 1.[25, 26] An answer to the question of the reliability of the power-law exponents in ref 1 would require quantitative analysis of both the true and sizing error distributions of lipid nanoparticle diameters. However, the susceptibility of such exponents to suppression by sizing errors of the standard deviation presents a surprising possibility—observation of an expected exponent, which would ordinarily build confidence, could instead result from a biased measurement of an unexpectedly high exponent. For this reason, accurate measurements, rather than hypothetical expectations of fluorophore loading trends that neglect the full scope of interactions of light with dielectric nanostructures, are necessary to support reliable studies of lipid nanoparticles for vaccine and medicine delivery, as well as studies that involve nanoplastic standards for hazard assessment. Even so, in these[25, 26] and other[29, 30] studies, any sizing error is of concern. This Letter to the Editor not only elucidates this problem but also proposes a potential solution by the deconvolution of experimental histograms with reference histograms to inform measurement error models.[23] Future studies may benefit from the use of such models to avoid biases in property correlation and parameter estimation.

## AUTHOR INFORMATION

**Corresponding Author**
    **Samuel M. Stavis** – Microsystems and Nanotechnology Division, National Institute of Standards and Technology, Gaithersburg, Maryland 20899, United States

**Authors**
    **Andrew C. Madison** – Microsystems and Nanotechnology Division, National Institute of Standards and Technology, Gaithersburg, Maryland 20899, United States
    **Adam L. Pintar** – Statistical Engineering Division, National Institute of Standards and Technology, Gaithersburg, Maryland 20899, United States
    **Craig R. Copeland** – Microsystems and Nanotechnology Division, National Institute of Standards and Technology, Gaithersburg, Maryland 20899, United States
    **Natalia Farkas** – Microsystems and Nanotechnology Division, National Institute of Standards and Technology, Gaithersburg, Maryland 20899, United States; Theiss Research, La Jolla, California 92037, United States; Building and Fire Sciences, United States Forest Service, Forest Products Laboratory, Madison, Wisconsin 53726, United States

**Author Contributions**
‡Equal contribution

## ACKNOWLEDGMENTS

The authors gratefully acknowledge an anonymous reviewer, J. Fagan, J. A. Liddle, J. Kramar, and E. A. Strychalski for helpful comments, and the National Institute of Standards and Technology Strategic and Emerging Research Initiative in support of the Circular Economy Program.

## DISCLAIMER

The identification of a commercial product in Ref. 19 is for specification only and does not imply recommendation.

## ASSOCIATED CONTENT

**Supporting Information**
Supporting Information is available free of charge at https://pubs.acs.org/doi/(...).

    Terminology; arithmetic and harmonic means; diameter histograms; histogram digitization; digitization uncertainty; bin data; error modeling and simulation; computational deconvolution; simulation parameters; slope suppression; average effects (PDF)

**Preprint**
This Letter to the Editor has an associated preprint:

    Madison, A. C.; Pintar, A. L.; Copeland, C. R.; Farkas, N.; Stavis, S. M. Letter to the Editor Concerning Simultaneous, Single-Particle Measurements of Size and Loading Give Insights into the Structure of Drug-Delivery Nanoparticles. 2023, arXiv:2302.01778. *arXiv*. https://arxiv.org/abs/2302.01778 (accessed March 20, 2023).

# Supporting Information

**Letter to the Editor Concerning "Simultaneous, Single-Particle Measurements of Size and Loading Give Insights into the Structure of Drug-Delivery Nanoparticles"**

Andrew C. Madison,[†,⊥] Adam L. Pintar,[‡,⊥] Craig R. Copeland,[†] Natalia Farkas,[†,§,∥] and Samuel M. Stavis[*,†]

## Index

[†]Microsystems and Nanotechnology Division, National Institute of Standards and Technology, Gaithersburg, Maryland 20899, United States. [⊥]Equal contributions. [‡]Statistical Engineering Division, National Institute of Standards and Technology, Gaithersburg, Maryland 20899, United States. [§]Theiss Research, La Jolla, California 92037, United States. [∥]Building and Fire Sciences, United States Forest Service, Forest Products Laboratory, Madison, Wisconsin 53726, United States. *Email: sstavis@nist.gov.

## Table S1. Terminology

| term | definition |
|---|---|
| $d_h$ | Hydrodynamic diameter of a single nanoparticle |
| $k_B$ | Boltzmann constant |
| $T$ | Absolute temperature of the dispersion fluid |
| $\eta$ | Kinematic viscosity of the dispersion fluid |
| $C$ | Hydrodynamic correction of nanoparticle diffusivity |
| $D_{\parallel}$ | Diffusivity parallel to confining slit surfaces |
| $d_s$ | Slit depth or microwell depth |
| $d_{h_i}$ | Bin center $i$ of hydrodynamic diameter histogram |
| $w_i$ | Bin weight $i$ of hydrodynamic diameter histogram |
| $\bar{d}_{h_H}$ | Harmonic mean of hydrodynamic diameter distribution |
| $\bar{d}_{h_A}$ | Arithmetic mean of hydrodynamic diameter distribution |
| $\sigma_{d_h}$ | Standard deviation of hydrodynamic diameter distribution |
| $\mathcal{N}$ | Normal distribution |
| $\mathcal{M}$ | Multinomial distribution |
| $S_{d_j}$ | Random variable representing diameter scale factor of simulation $j$ |
| $\mu_{S_d}$ | Mean scale factor of diameter axis of hydrodynamic diameter histogram |
| $\epsilon_{S_d}$ | Relative standard error of mean scale factor of hydrodynamic diameter histogram |
| $B_{d_j}$ | Random variable representing diameter bin width of simulation $j$ |
| $\mu_{\delta_d}$ | Mean bin width of hydrodynamic diameter histogram |
| $\epsilon_{\delta_d}$ | Standard error of the mean bin width of hydrodynamic diameter histogram |
| $d_{i,j}$ | Random variable representing the central diameter of bin $i$ and simulation $j$ of hydrodynamic diameter histogram |
| $d_{\xi_i}$ | Left edge of bin $i$ of hydrodynamic diameter histogram |
| $\sigma_{\xi_d}$ | Standard deviation of left edge of bin value, corresponding to diameter, of hydrodynamic diameter histogram |
| $S_{w_j}$ | Random variable representing weight scale factor of simulation $j$ of hydrodynamic diameter histogram |
| $\mu_{s_w}$ | Mean scale factor of weight axis of hydrodynamic diameter histogram |
| $\epsilon_{S_w}$ | Relative standard error of mean scale factor of weight axis of hydrodynamic diameter histogram |
| $\sigma_{\xi_w}$ | Standard deviation of top edge of bin value, corresponding to weight, of hydrodynamic diameter histogram |
| $\boldsymbol{W}_j$ | Random vector of weights of simulation $j$ of hydrodynamic diameter histogram |
| $W_{i,j}$ | Components of $\boldsymbol{W}_j$ of hydrodynamic diameter histogram |
| $\boldsymbol{w}_\xi$ | Mean vector of weights from the top edges of bins of hydrodynamic diameter histogram |
| $w_{\xi_i}$ | Components of $\boldsymbol{w}_\xi$ from the top edges of bins of hydrodynamic diameter histogram |
| $d$ | Independent variable of diameter, setting aside the divergence of hydrodynamic and steric sizes |
| $d_{\text{ref}}$ | Random variable representing the reference diameter distribution |
| $\epsilon_d$ | Random variable representing an error distribution of diameter measurements |
| $M_d$ | Random variable representing a measurement distribution of diameters |
| $\mathcal{LN}$ | Lognormal distribution |
| $\mu_{d_{\text{ref}}}$ | Mean of the reference diameter distribution |
| $\sigma_{d_{\text{ref}}}$ | Standard deviation of the reference diameter distribution |
| $\mu_{\text{ref}}$ | Mean of the natural logarithm of the reference diameter distribution |
| $\sigma_{\text{ref}}$ | Standard deviation of the natural logarithm of the reference diameter distribution |
| $\mathcal{SN}$ | Skew-normal distribution |
| $\epsilon_{\mu_d}$ | Error of the mean of the reference diameter distribution |
| $\epsilon_{\sigma_d}$ | Error of the standard deviation of the reference diameter distribution |
| $\xi$ | Location parameter of the skew-normal distribution of diameter errors |
| $\omega$ | Scale parameter of the skew-normal distribution of diameter errors |
| $\alpha$ | Shape parameter of the skew-normal distribution of diameter errors |
| $\delta$ | Simplifying expression of the shape parameter of the skew-normal distribution of diameter errors |
| $r$ | Independent variable of radius, $d/2$, setting aside the divergence of hydrodynamic and steric sizes |
| $r_{\text{ref}}$ | Random variable representing the reference radius distribution |
| $I$ | Fluorescence intensity and dependent variable of power-law model |
| $\gamma$ | True scaling coefficient of power-law model |
| $\ln(\gamma)$ | True intercept of linear model of $\ln(\text{intensity})$ versus $\ln(\text{radius})$ |
| $\beta$ | True slope of linear model of $\ln(\text{intensity})$ versus $\ln(\text{radius})$ and true exponent of power-law model of intensity versus radius |
| $\epsilon_I$ | Random variable representing errors of $\ln(\text{intensity})$ |
| $\sigma_I$ | Standard deviation of errors of $\ln(\text{intensity})$ |
| $\beta_{\text{fit}}$ | Apparent $\beta$ for fits of the linear model to erroneous $\ln(\text{intensity})$ versus $\ln(\text{radius})$ data |
| $\ln(\gamma)_{\text{fit}}$ | Apparent $\ln(\gamma)$ for fits of the linear model to erroneous $\ln(\text{intensity})$ versus $\ln(\text{radius})$ data |
| $\epsilon_\beta$ | Errors of $\beta$ due to measurement errors of radius and intensity, $\beta_{\text{fit}} - \beta$ |
| $\epsilon_{\ln(\gamma)}$ | Errors of $\ln(\gamma)$ due to measurement errors of radius and intensity, $\ln(\gamma)_{\text{fit}} - \ln(\gamma)$ |

**Note S1. Arithmetic and harmonic means**
For a colloidal nanoparticle in a confining slit, a variant of the Stokes–Einstein–Sutherland equation expresses the inverse proportionality between diffusivity parallel to the slit surfaces and hydrodynamic diameter:

$$D_{||} = \frac{k_B}{3\pi} \cdot \frac{T}{\eta} \cdot \frac{C}{d_h} \tag{S1}$$

A distributional validation of a method to size single nanoparticles using eq S1 would require agreement of a sample diameter distribution with two reference values of the nanoparticle standard—an arithmetic mean and a standard deviation. The most straightforward approach to make this comparison would be to convert diffusivity to diameter for single nanoparticles and calculate the arithmetic mean and standard deviation of the sample diameters.

Implicit in ref 1 is an alternate calculation of the arithmetic mean of $D_{||}$ for $N$ single nanoparticles:

$$\overline{D}_{||_A} = \frac{1}{N} \sum_{i=1}^{N} D_{||_i} \tag{S2}$$

where $\overline{(\ldots)}_A$ denotes an arithmetic mean. Substituting eq S1 into eq S2 and solving for the corresponding mean diameter yields:

$$\bar{d}_h = \frac{k_B}{3\pi} \cdot \frac{T}{\eta} \cdot \frac{C}{\overline{D}_{||_A}} = \frac{N}{\sum_{i=1}^{N} \frac{1}{d_{h_i}}} \tag{S3}$$

The corresponding mean diameter is the harmonic mean diameter, which can differ from the arithmetic mean diameter:

$$\frac{N}{\sum_{i=1}^{N} \frac{1}{d_{h_i}}} = \bar{d}_{h_H} \not\equiv \bar{d}_{h_A} = \frac{1}{N} \sum_{i=1}^{N} d_{h_i} \tag{S4}$$

where $\overline{(\ldots)}_H$ denotes a harmonic mean. Without access to a reference diameter histogram, the harmonic mean diameter lacks a certain value for comparison and validation.

**Note S2. Diameter histograms**
Access to the data underlying the diameter histograms of Figure 1F of ref 1 would enable direct calculation of the harmonic mean, arithmetic mean, and standard deviation of the nanoparticle sample diameters. Lacking access to the underlying data, independent analysis of the histogram plots provides close approximations of these three values. Digitization of the histogram plots, using software[2] that previous studies have validated,[3] allows extraction of the bin data (Note S3 of this Letter). Each histogram has $N$ bins with index $i$. The center of each bin is a hydrodynamic diameter, $d_{h_i}$, in units of nanometers. The top edge of each bin is a weight, $w_i$, in units of counts, which can be a non-integer due to the averaging of histogram bins from at least two to three experiments. For each diameter histogram, the harmonic mean is:

$$\bar{d}_{h_H} = \frac{\sum_{i=1}^{N} w_i}{\sum_{i=1}^{N} \frac{w_i}{d_{h_i}}} \tag{S5}$$

the arithmetic mean is:

$$\bar{d}_{h_A} = \frac{\sum_{i=1}^{N} w_i d_{h_i}}{\sum_{i=1}^{N} w_i} \tag{S6}$$

and the standard deviation is:

$$\sigma_{d_h} = \sqrt{\frac{\sum_{i=1}^{N} w_i (d_{h_i} - \bar{d}_{h_A})^2}{\frac{N'-1}{N'} \sum_{i=1}^{N} w_i}} \tag{S7}$$

where $N'$ is the number of bins with non-zero weights.

**Note S3. Histogram digitization**

The extraction of bin data from histogram plots is a microcosm of a measurement involving image analysis and requiring uncertainty evaluation. Errors of diameter and weight result from digitizing the diameter histogram plots, while intrinsic variability of weight results from finite sampling of the diameter distributions. Monte-Carlo simulations propagate input distributions from digitizing error and sampling variability to output distributions of harmonic means, arithmetic means, and standard deviations. Normal distributions approximate digitization errors and the multinomial distribution accounts for sampling variability. The output distributions manifest measurement uncertainty and are approximately normal, having standard deviations that correspond to the 68% coverage intervals in Table 1 of the main text of this Letter.

This analysis begins with the calibration of scale factors. For each axis of each histogram plot, at least 30 replicate measurements of the distance between two tick marks yields a ratio of axis units to plot pixels, providing a mean and standard deviation of scale factor for each axis. The relative standard errors of the mean scale factors provide relative uncertainties of the scale conversion that occurs within the digitizing software.[2] The diameter scale factor and relative uncertainty are the same for the five histogram plots, whereas the weight scale factor and relative uncertainty differ for each histogram plot (Table S2 of this Letter).

**Table S2. Digitization uncertainty**

| microwell depth $d_s$ (nm) | diameter scale-factor uncertainty $\epsilon_{S_d}$ (%) | diameter bin-width uncertainty $\epsilon_{\delta_d}$ (nm) | diameter single-bin uncertainty $\sigma_{\xi_d}$ (nm) | weight scale-factor uncertainty $\epsilon_{S_w}$ (%) | weight single-bin uncertainty $\sigma_{\xi_w}$ (counts) |
|---|---|---|---|---|---|
| 200 | 0.018 | 0.03 | 0.10 | 0.16 | 0.08 |
| 350 | 0.018 | 0.03 | 0.10 | 0.22 | 0.53 |
| 500 | 0.018 | 0.03 | 0.10 | 0.16 | 0.75 |
| 800 | 0.018 | 0.03 | 0.10 | 0.16 | 0.29 |
| 1200 | 0.018 | 0.03 | 0.10 | 0.24 | 0.19 |

The analysis continues to the measurement of single bins. For each non-zero bin that is visible outside of the inset plot in Figure 1F, extraction of the plot coordinates at the corner of the left and top edges of the bin yields a diameter and weight (Tables S3–S7 of this Letter). Subtraction of the left edges of adjacent bins yields a bin width with an arithmetic mean of 5.0 nm and a standard error of 0.03 nm. The left edges of the bins evidently correspond to nominal values of 10.0, 15.0, 20.0 nm, and so on. Addition of half of the bin width to the left edges of the bins yields the center positions of the bins, which lack data markers but are near to the nominal values of 12.5, 17.5, 22.5 nm, and so on. To evaluate diameter uncertainties, at least 30 replicate measurements along the vertical weight axis, with a diameter coordinate of 0 nm, yield a standard deviation of 0.10 nm. To evaluate weight uncertainties, at least 30 replicate measurements along the horizontal diameter axis, with a weight coordinate of 0 counts, yield a standard deviation ranging from 0.08 to 0.75 counts, depending on the weight scale factor (Table S2 of this Letter).

To evaluate sampling variability, a bootstrap algorithm resamples the histograms by drawing replicate bin weights from a multinomial distribution. This distribution models the random sampling of a nanoparticle from a histogram with varying bin probabilities and for a constant number of samples. The probability for each bin is the ratio of the bin weight resulting from the digitization of the plot data to the sum of all bin weights from the histogram. The sum of all bin weights from the histogram, rounded to the nearest integer, defines the sample size.

The following equations summarize the sampling process of $2.5\times10^4$ Monte-Carlo simulations for each diameter histogram plot of Figure 1F of ref 1. These simulations involve the normal distribution $\mathcal{N}$ and the multinomial distribution $\mathcal{M}$. The $j^{\text{th}}$ simulation of the centers of diameter bins $d_{\text{h}_{i,j}}$ is:

$$S_{d_j} \sim \mathcal{N}(\mu_{S_d}, \epsilon_{S_d}^2) \tag{S8}$$

$$B_{d_j} \sim \mathcal{N}(\mu_{\delta_d}, \epsilon_{\delta_d}^2) \tag{S9}$$

$$d_{i,j}|B_{d_j} \sim \mathcal{N}(d_{\xi_i} + \frac{B_{d_j}}{2}, \sigma_{\xi_d}^2) \tag{S10}$$

$$d_{\text{h}_{i,j}} = S_{d_j} d_{i,j} \tag{S11}$$

where $S_{d_j}$ is a random variable representing the diameter scale factor, which has an arithmetic mean of $\mu_{s_d} = 1$ after normalization and a standard deviation equal to a relative standard error of $\epsilon_{s_d} = 0.018\%$; $B_{d_j}$ is a random variable representing the diameter bin width, which has an arithmetic mean of $\mu_{\delta_d} = 5.0$ nm and a standard deviation equal to $\epsilon_{\delta_d} = 0.03$ nm; and $d_{i,j}$, which is conditional on $B_{d_j}$, is a random variable representing the central diameter of bin $i$, which has an arithmetic mean equal to the sum of the left edge of the bin $d_{\xi_i}$ and half the bin width, and a standard deviation of single-bin measurements of diameter of $\sigma_{\xi_d} = 0.10$ nm. The $j^{\text{th}}$ simulation of the vector of weights of the diameter bins $\boldsymbol{w}_j$ is:

$$S_{w_j} \sim \mathcal{N}(\mu_{s_w}, \epsilon_{s_w}^2) \tag{S12}$$

$$W_{i,j} \sim \mathcal{N}(w_{\xi_i}, \sigma_{\xi_w}^2) \tag{S13}$$

$$\boldsymbol{W}_j = \begin{pmatrix} W_{1,j} \\ \vdots \\ W_{N,j} \end{pmatrix} \tag{S14}$$

$$\boldsymbol{w}_j | S_{w_j}, \boldsymbol{W}_j \sim \mathcal{M}\left(S_{w_j} \sum_{i=1}^{N} W_{i,j}, \frac{S_{w_j} W_j}{S_{w_j} \sum_{i=1}^{N} W_{i,j}}\right) \tag{S15}$$

$$\boldsymbol{w}_j = \begin{pmatrix} w_{1,j} \\ \vdots \\ w_{N,j} \end{pmatrix} \tag{S16}$$

where $S_{w_j}$ is a random variable representing the weight scale factor, which has an arithmetic mean of $\mu_{s_w} = 1$ after normalization, and a standard deviation equal to a relative standard error $\epsilon_{s_w}$ ranging from 0.16 to 0.24%; and $\boldsymbol{W}_j$ is a random vector of weights, which has components $W_{i,j}$, an arithmetic mean $\boldsymbol{w}_\xi$ from the vector of the top edges of all of the bins, and a standard deviation of single-bin measurements of weight $\sigma_{\xi_w}$ ranging from 0.08 to 0.75 counts. In this way, eqs S11 and S16 resample the diameters and weights, respectively, of the histogram plots.

**Table S3. Bin data for a microwell depth of 200 nm**

| diameter, $d_{\xi_i}$ (nm) | weight, $w_{\xi_i}$ (counts) |
|---|---|
| 20.1 | 0.5 |
| 25.1 | 3.9 |
| 30.0 | 5.3 |
| 35.1 | 3.0 |
| 40.1 | 4.8 |
| 45.0 | 9.0 |
| 50.0 | 10.5 |
| 55.0 | 11.4 |
| 60.1 | 7.5 |
| 65.0 | 14.4 |
| 70.1 | 8.9 |
| 75.0 | 6.4 |
| 80.0 | 7.5 |
| 85.1 | 4.5 |
| 90.1 | 4.0 |
| 95.0 | 2.3 |
| 100.0 | 1.8 |
| 105.0 | 3.3 |
| 110.1 | 1.0 |
| 115.0 | 2.3 |
| 120.1 | 1.1 |
| 130.0 | 0.5 |
| 135.0 | 0.8 |
| 140.0 | 0.5 |
| 145.0 | 2.3 |
| 155.0 | 0.5 |
| 160.0 | 0.8 |
| 170.0 | 0.6 |
| 175.1 | 0.9 |
| 185.0 | 0.8 |
| 205.0 | 0.5 |
| 215.1 | 0.6 |

**Table S4. Bin data for a microwell depth of 350 nm**

| diameter, $d_{\xi_i}$ (nm) | weight, $w_{\xi_i}$ (counts) |
|---|---|
| 10.1 | 3.3 |
| 15.1 | 4.3 |
| 20.1 | 19.3 |
| 25.0 | 47.5 |
| 30.0 | 81.9 |
| 35.0 | 96.9 |
| 40.0 | 124.1 |
| 45.1 | 123.6 |
| 50.1 | 109.5 |
| 55.1 | 99.3 |
| 60.1 | 81.3 |
| 65.0 | 65.7 |
| 70.0 | 42.4 |
| 75.0 | 30.7 |
| 80.0 | 23.4 |
| 85.0 | 16.6 |
| 90.1 | 12.2 |
| 95.1 | 9.2 |
| 100.1 | 8.7 |
| 105.0 | 8.7 |
| 110.0 | 4.8 |

**Table S5. Bin data for a microwell depth of 500 nm**

| diameter, $d_{\xi_i}$ (nm) | weight, $w_{\xi_i}$ (counts) |
|---|---|
| 20.0 | 7.8 |
| 25.1 | 24.9 |
| 30.0 | 84.9 |
| 35.0 | 139.2 |
| 40.0 | 154.1 |
| 45.0 | 184.8 |
| 50.0 | 150.5 |
| 55.5 | 132.7 |
| 60.0 | 101.9 |
| 65.0 | 59.1 |
| 70.1 | 53.3 |
| 75.1 | 38.3 |
| 80.0 | 30.4 |
| 85.0 | 19.7 |
| 90.0 | 6.8 |
| 95.0 | 11.1 |
| 100.0 | 5.4 |
| 105.1 | 3.2 |

**Table S6. Bin data for a microwell depth of 800 nm**

| diameter, $d_{\xi_i}$ (nm) | weight, $w_{\xi_i}$ (counts) |
|---|---|
| 15.0 | 2.9 |
| 20.0 | 5.8 |
| 25.1 | 13.4 |
| 30.0 | 33.9 |
| 35.1 | 69.7 |
| 40.1 | 78.3 |
| 45.0 | 88.5 |
| 50.0 | 93.9 |
| 55.0 | 67.5 |
| 60.2 | 55.2 |
| 65.1 | 39.0 |
| 70.1 | 26.4 |
| 75.1 | 18.8 |
| 80.1 | 10.1 |
| 85.1 | 5.8 |
| 90.1 | 3.2 |
| 95.0 | 2.5 |
| 100.0 | 1.1 |
| 105.0 | 1.1 |

**Table S7. Bin data for a microwell depth of 1200 nm**

| diameter, $d_{\xi_i}$ (nm) | weight, $w_{\xi_i}$ (counts) |
|---|---|
| 15.0 | 2.4 |
| 20.0 | 7.7 |
| 25.1 | 26.5 |
| 30.0 | 44.0 |
| 35.0 | 58.8 |
| 40.1 | 68.6 |
| 45.1 | 53.0 |
| 50.1 | 45.0 |
| 55.1 | 45.6 |
| 60.1 | 24.4 |
| 65.0 | 18.1 |
| 70.0 | 19.1 |
| 75.0 | 14.4 |
| 80.0 | 10.4 |
| 85.0 | 6.5 |
| 90.1 | 6.7 |
| 95.0 | 4.4 |
| 100.0 | 3.6 |
| 110.0 | 3.6 |

**Note S4. Error modeling and simulation**

Monte-Carlo simulations elucidate the effects of the measurement errors that are relatively certain in ref 1 on intensity–size trends. This analysis sets aside the divergence of measurements of hydrodynamic diameter by convex lens-induced confinement microscopy and of steric diameter by transmission electron microscopy. This simplification also occurs in ref 1, separately from the ensemble measurements by dynamic light scattering therein, and allows a quantitative study of measurement errors.

Starting with diameter errors, random and systematic effects are evident in the benchmark measurements of the nanoparticle standard in ref 1 (Figure 1a of this Letter). A statistical model equates the measurement result for the apparent diameter of a single nanoparticle, $M_d$, to the sum of the true diameter, $d_{\text{ref}}$, which is a random variable representing the reference diameter distribution, and a measurement error, $\epsilon_d$, which is a random variable representing an error distribution:

$$M_d = d_{\text{ref}} + \epsilon_d \tag{S17}$$

The lognormal distribution, $\mathcal{LN}$, provides a characteristic approximation of the reference diameter distribution. The arithmetic mean, $\mu_{d_{\text{ref}}}$, and variance, $\sigma^2_{d_{\text{ref}}}$, of the reference diameter distribution parameterize the mean, $\mu_{\text{ref}}$, and variance, $\sigma^2_{\text{ref}}$, of the lognormal distribution:

$$d_{\text{ref}} \sim \mathcal{LN}(\mu_{\text{ref}}, \sigma^2_{\text{ref}}) \tag{S18}$$

$$\mu_{\text{ref}} = \ln\left(\frac{\mu^2_{d_{\text{ref}}}}{\sqrt{\mu^2_{d_{\text{ref}}} + \sigma^2_{d_{\text{ref}}}}}\right) \tag{S19}$$

$$\sigma^2_{\text{ref}} = \ln\left(1 + \frac{\sigma^2_{d_{\text{ref}}}}{\mu^2_{d_{\text{ref}}}}\right) \tag{S20}$$

The skew-normal distribution, $\mathcal{SN}$, provides an empirical approximation of the error distributions of the diameter measurements in ref 1. The mean of an error distribution, $\epsilon_{\mu_d}$, is the difference between the mean of the reference diameter distribution and the arithmetic mean of a diameter histogram of Figure 1F of ref 1. The variance of an error distribution is the difference between a variance of a diameter histogram in Figure 1F in ref 1, $(\sigma_{d_{\text{ref}}} + \epsilon_{\sigma_d})^2$, and the variance of the reference diameter distribution, $\sigma^2_{d_{\text{ref}}}$, where $\epsilon_{\sigma_d}$ is the difference between the standard deviation of the reference diameter distribution and the standard deviation of a diameter histogram of Figure 1f of ref 1. In addition to the diameter distribution errors $\epsilon_{\mu_d}$ and $\epsilon_{\sigma_d}$, the shape parameter, $\alpha$, influences both the location parameter, $\xi$, and scale parameter, $\omega$, of an error distribution:

$$\epsilon_d \sim \mathcal{SN}(\xi, \omega^2, \alpha) \tag{S21}$$

$$\delta = \frac{\alpha}{\sqrt{1+\alpha}} \tag{S22}$$

$$\xi = \epsilon_{\mu_d} - \omega\delta\sqrt{\frac{2}{\pi}} \tag{S23}$$

$$\omega = \sqrt{\frac{(\sigma_{d_{\text{ref}}} + \epsilon_{\sigma_d})^2 - \sigma^2_{d_{\text{ref}}}}{1 - \frac{2\delta^2}{\pi}}} \tag{S24}$$

Computational deconvolution of the diameter histograms in Figure 1f of ref 1 with the reference diameter distribution assumes certain values of $\epsilon_{\mu_d}$ and $\epsilon_{\sigma_d}$ and adjusts $\alpha$ to minimize the Kolmogorov–Smirnov distance between the empirical distribution functions from ref 1 and the empirical distribution functions from Monte-Carlo simulations of $M_d$. The resulting values of $\alpha$ range from 9.5 to 12.5 (Figure S1 of this Letter). An alternative selection of a normal distribution, rather than a lognormal distribution, of reference diameters $d_{\text{ref}}$ for the nanoparticle standard increases the arithmetic mean of $\alpha$ only slightly from 10.5 to 10.7, with the remainder of the main results remaining approximately the same (not shown).

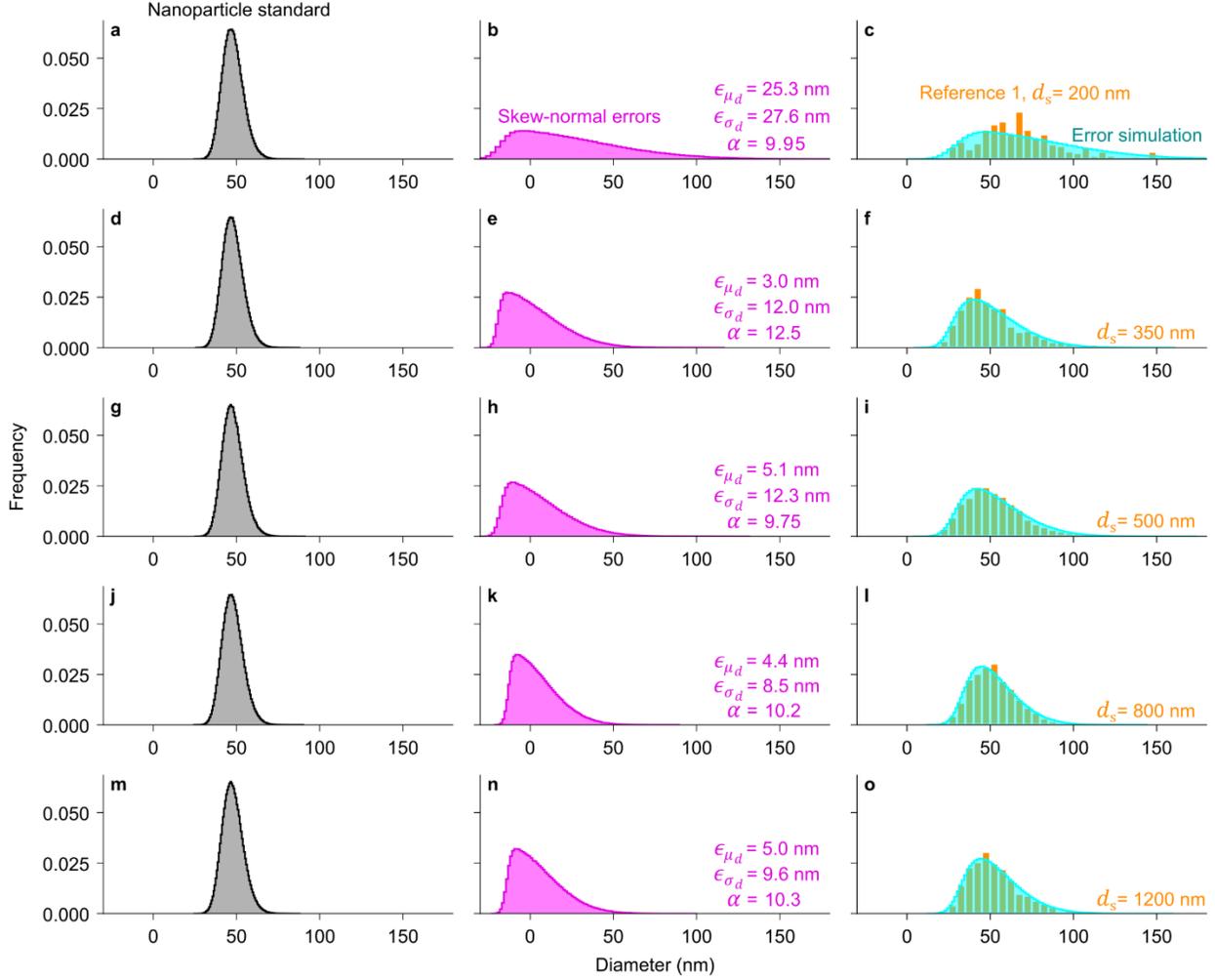

**Figure S1. Computational deconvolution.** (a, d, g, j, m) Histograms showing the (gray) diameter distributions of the nanoparticle standard with the reference values parameterizing a lognormal distribution. (b, e, h, k, n) Histograms showing (magenta) skew-normal error distributions resulting from computational deconvolution. (c, f, i, l, o) Histograms showing (orange) experimental results for the diameter measurements in ref 1 and (cyan) simulation results from convolution of the lognormal reference diameter distributions in (a, d, g, j, m) with the skew-normal error distributions in (b, e, h, k, n). Histograms correspond to microwell depths of (a-c) 200 nm, (d-f) 350 nm, (g-i) 500 nm, (j-l) 800 nm, and (m-o) 1200 nm. For clear visualization, histograms of the nanoparticle standard, error distributions, and simulations consist of 1×10$^6$ samples. The orange histogram in (c) shows significant sampling variability from a relatively low count of single nanoparticles in ref 1.

To calculate a true intensity–radius trend, changing the independent variable from diameter to radius as in ref 1, a power law constrains intensity, $I$, to depend on reference radius, $r_{\text{ref}}$, by the model, $I = \gamma r_{\text{ref}}^{\beta}$, with a coefficient, $\gamma$, and an exponent, $\beta$. After transformation by the natural logarithm, the corresponding linear model has a slope, $\beta$, and intercept, $\ln(\gamma)$.

Unknown combinations of measurement variability and sample heterogeneity are evident in intensity measurements of lipid nanoparticles in Figures 2, 3, and 6 of ref 1. Building on the power-law model in the main text of this Letter, a normal distribution, $\mathcal{N}$, provides an empirical approximation of the distribution of $\ln(\text{intensity})$ errors with an arithmetic mean of zero and a corresponding range of variance, $\sigma_I^2$:

$$\ln(I) = \beta \ln(r_{\text{ref}}) + \ln(\gamma) + \epsilon_I \tag{S25}$$

$$\epsilon_I \sim \mathcal{N}(0, \sigma_I^2) \tag{S26}$$

Ordinary least-squares fits of linear models to the synthetic measurement results for the independent variable of $\ln(\text{radius})$, $\ln r = \ln[\mathcal{M}_d/2]$, versus the dependent variable of $\ln(\text{intensity})$, yield apparent values of the slope, $\beta_{\text{fit}}$, and intercept, $\ln(\gamma)_{\text{fit}}$, which may differ from the true values in each simulation. The simulations yield errors of the slope, $\epsilon_\beta$, and intercept, $\epsilon_{\ln(\gamma)}$, as the differences between the parameters in fits of models and the corresponding true values of input parameters:

$$\epsilon_\beta = \beta_{\text{fit}} - \beta \tag{S27}$$

$$\epsilon_{\ln(\gamma)} = \ln(\gamma)_{\text{fit}} - \ln(\gamma) \tag{S28}$$

To implement a factorial study, each simulation takes six inputs, in addition to the arithmetic mean and standard deviation of the reference diameter distribution, and estimates the errors of the fit parameters for a linear model of $\ln(\text{intensity})$ versus $\ln(\text{radius})$ (Table S8 of this Letter). Repeating the simulation 100 times for each combination of input parameters samples potential errors of the fit parameters. Grouping the resulting error estimates by the unique values of a single input parameter, and averaging over the values of the other input parameters, isolates the average effect of each input parameter on the error of the fit parameters (Figure S3 of this Letter).

For the reference diameters and sizing errors in the effort to validate, the error of the standard deviation of nanoparticle diameters has the largest average effect (Figure S3b,h of this Letter). This result is consistent with the canonical example of slope suppression in linear regression, which motivates the development of a measurement error model to improve parameter estimation by linear regression.[4] Other simulation parameters have lesser or no average effects. Both the true slope of the linear model of $\ln(\text{intensity})$ versus $\ln(\text{radius})$ and the error of the mean of nanoparticle diameters have lesser average effects on the error of the fit parameters. The shape parameter of the skew-normal diameter error distribution, the true intercept of the linear model of $\ln(\text{intensity})$ versus $\ln(\text{radius})$, and the standard deviation of $\ln(\text{intensity})$ have no effects on the error of the fit parameters (Figure S3 of this Letter).

**Table S8. Simulation parameters** [a]

| symbol | definition | minimum value | maximum value | unit | number of values |
|---|---|---|---|---|---|
| **$\mu_{d_{\text{ref}}}$** | **Mean of the reference diameter distribution** | | 47.8 | nm | 1 |
| **$\sigma_{\text{ref}}$** | **Standard deviation of the reference diameter distribution** | | 6.3 | nm | 1 |
| $\epsilon_{\mu_d}$ | Error of the arithmetic mean of the diameter distribution | 0 | 35 | nm | 20 |
| $\epsilon_{\sigma_d}$ | Error of the standard deviation of the diameter distribution | 0 | 35 | nm | 20 |
| $\alpha$ | Shape parameter of the skew-normal diameter error distribution | 9 | 13 | – | 5 |
| $\beta$ | True slope of the linear model of $\ln(\text{intensity})$ versus $\ln(\text{radius})$ | 1.5 | 3.5 | $\ln(\text{nm})^{-1}$ | 7 |
| $\ln(\gamma)$ | True intercept of the linear model of $\ln(\text{intensity})$ versus $\ln(\text{radius})$ | -4.5 | -2.5 | $\ln(\text{arb.})$ | 5 |
| $\sigma_I$ | Standard deviation of $\ln(\text{intensity})$ | 0.3 | 0.7 | $\ln(\text{arb.})$ | 5 |

[a] Bold text indicates reference values that remain constant in the simulation. The sample size of each simulation is 500 nanoparticles. Each combination of input parameters occurs in replicates of 100 for a total of $3.5 \times 10^7$ simulations in a full factorial analysis.

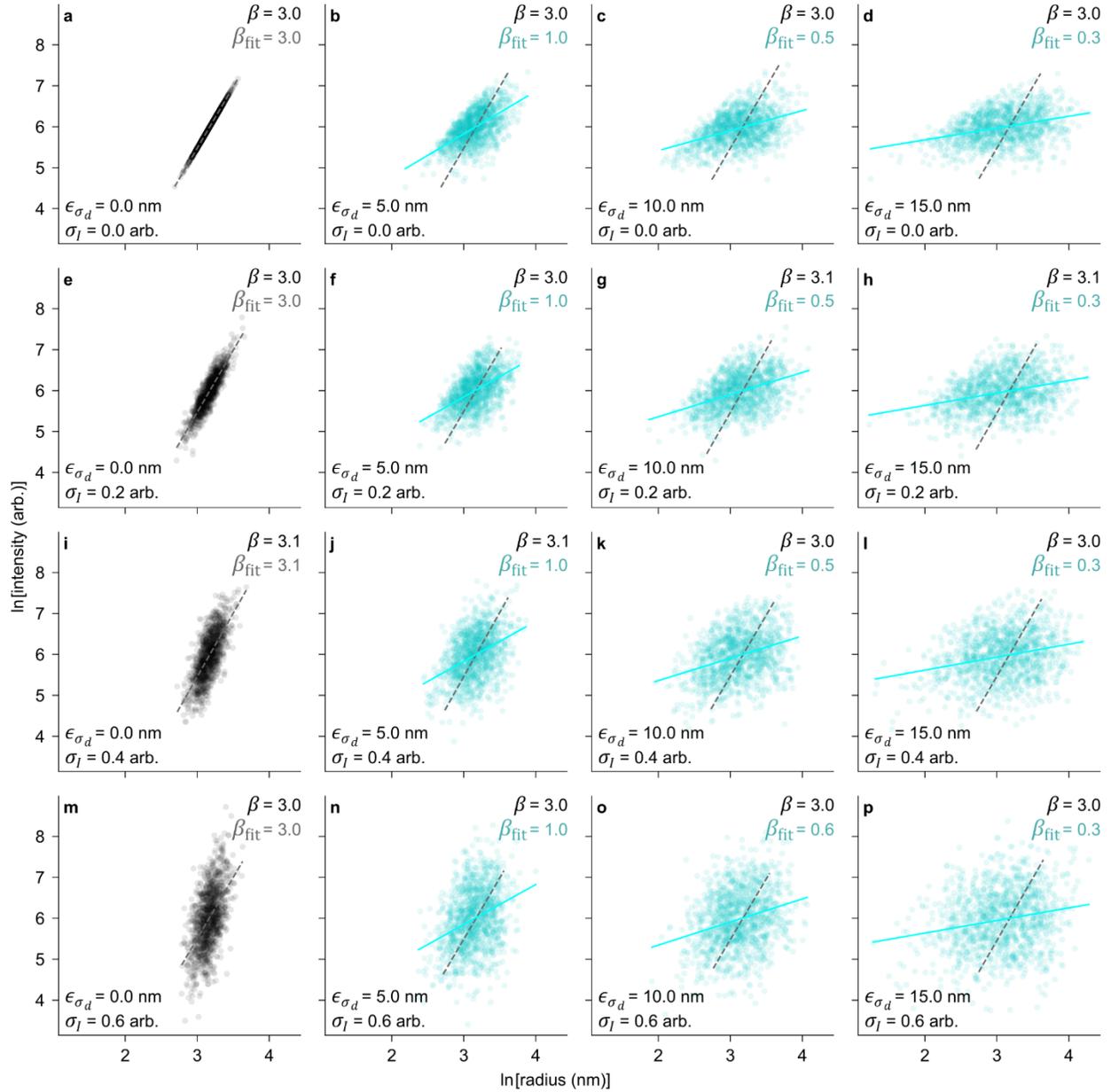

**Figure S2. Slope suppression.** (a-p) Scatter plots showing certain effects of errors of the diameter standard deviation and of intensity, $\epsilon_{\sigma_d}$ and $\sigma_I$, on fits of a power-law model to (gray) reference data and to (cyan) simulation data for a true value of $\beta = 3.0$. To clearly show the onset of slope suppression, the range of errors differs slightly from that of the simulations in Table S8. Representative fluctuations of true values of $\beta$ around 3.0 are due to sampling variability for 1208 nanoparticles, as in Figure 1c of this Letter.

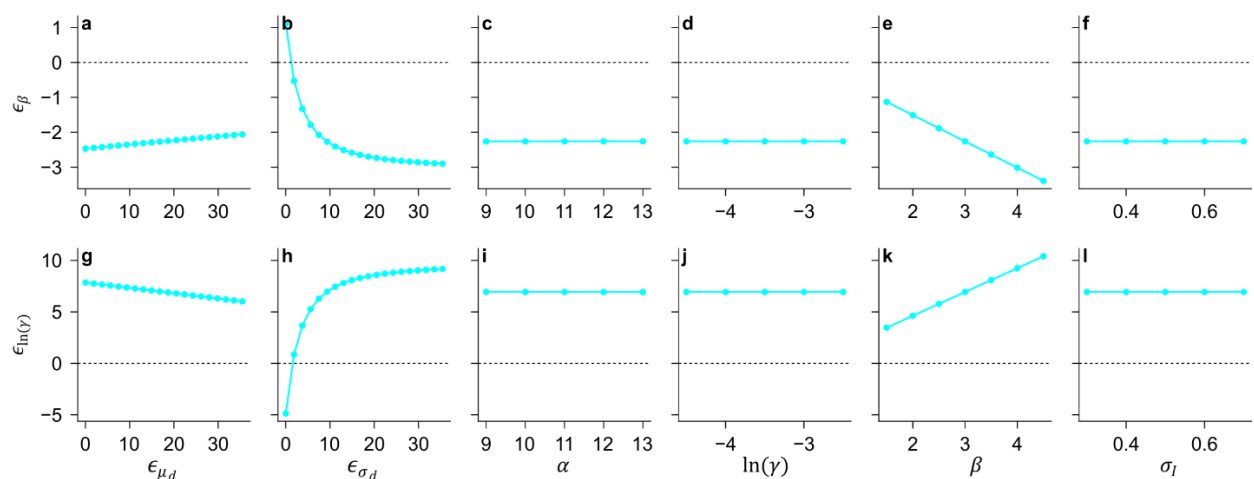

**Figure S3. Average effects.** Plots showing the average effects of simulation parameters on (a-f) error of the slope parameter and (g-l) error of the intercept parameter of the linear model of ln(intensity) versus ln(radius). The corresponding simulation parameters are in Table S8 of this Letter.

## SUPPORTING REFERENCES